\documentclass{article}
\usepackage{ijcai25}

\usepackage{times}
\usepackage{soul}
\usepackage{url}
\usepackage[hidelinks]{hyperref}
\usepackage[utf8]{inputenc}
\usepackage[small]{caption}
\usepackage{graphicx}
\usepackage{amsmath}
\usepackage{amsthm}
\usepackage{booktabs}
\usepackage{algorithm}
\usepackage{algorithmic}
\usepackage[switch]{lineno}

\title{RedTeamLLM: an Agentic AI framework for offensive security}

\author{
Brian Challita$^1$
\and
Pierre Parrend$^{1,2}$\and
\affiliations
$^1$Laboratoire de Recherche de l’EPITA, 14-16 Rue Voltaire, 94270 Le Kremlin-Bicêtre, France\\
$^2$ICube, UMR 7357, Université de Strasbourg, CNRS, 300 bd Sébastien Brant - CS 10413 - F-67412 Illkirch Cedex\\
\emails
\{brian.challita, pierre.parrend\}@epita.fr
}

\begin{document}

\maketitle

\begin{abstract}
    From automated intrusion testing to discovery of zero-day attacks before software launch, agentic AI calls for great promises in security engineering. This strong capability is bound with a similar threat: the security and research community must build up its models before the approach is leveraged by malicious actors for cybercrime. We therefore propose and evaluate RedTeamLLM, an integrated architecture with a comprehensive security model for automatization of pentest tasks. RedTeamLLM follows three key steps: summarizing, reasoning and act, which embed its operational capacity.
    This novel framework addresses four open challenges: plan correction, memory management, context window constraint, and generality vs. specialization. Evaluation is performed through the automated resolution of a range of entry-level, but not trivial, CTF challenges. The contribution of the reasoning capability of our agentic AI framework is specifically evaluated.
\end{abstract}


\section{Introduction}

The recent strengthening of Agentic AI \cite{hughes2025ai} approaches poses major challenges in the domains of cyberwarfare and geopolitics \cite{oesch2025agentic}. LLMs are already commonly used for cyber operations for augmenting human capabilities and automating common tasks \cite{yao2024survey,chowdhury2024breaking}. They already pose significant ethical and societal challenges \cite{malatji2024artificial}, and a great threat of proliferation of cyberdefence and -attack capabilities , which were so far only available for nation-state level actors. Whereas there current recognized capabilities are still bound to the rapid analysis of malicious code or rapid decision taking in alert triage, and they pose significant trust issues \cite{sun2024trustllm}, there expressivity and knowledge-base are rapidly ramping up.
In this context, Agentic AI, \textit{i.e.} autonomous AI systems that are capable of performing a set of complex tasks that span over long periods of time without human supervision \cite{acharya2025agentic}, is opening a brand new type of cyberthreat. They follow two complementary strategies: goal orientation, and reinforcement learning, which have the capability to dramatically accelerate the execution of highly technical operations, such as cybersecurity actions, while supporting a diversification of supported tasks.

In the defense landscape, cyberwarfare takes a singular position, and targets espionage, disruption, and degradation of information and operational systems of the adversary. More than in traditional arms, skill is a strong limiting factor, especially since targeting critical defense systems heavily relies on the exploitation of rare, unknown vulnerabilities, which are most often than not 0-days threats. Actually, whereas financial criminality aims at money extorsion and thus targets a broad range of potential victims to exploit the weakest ones, defense operations aim at entering and disrupting highly exposed, and highly protected, technical environments, where known vulnerabilities are closed very quickly. In this context, operational capability relies so far in talented analysts capable of discovering novel vulnerabilities. This high-skill, high-mean game could face a brutal end with the advent of tools capable of discovering new exploitable flows at the heart of the software, thus enabling smaller actors to exhibit a highly asymetric threats capable of disrupting critical infrastructures, or launching large-scale disinformation campaigns. Agentic AI has the capability to provide such a tool, and LLMs themselves in their stand-alone versions, have already proved capable of detecting these famous 0-day vulnerabilities: Microsoft has published, with the help of its Copilot tools, no less that 20 (!!) vulnerabilities in the Grub2, U-Boot and barebox bootloaders since late 2024 \footnote{https://www.microsoft.com/en-us/security/blog/2025/03/31/analyzing-open-source-bootloaders-finding-vulnerabilities-faster-with-ai/}.

This is the public side of the medal, by a company who seeks to advertise its software development environment, and create some noise on vulnerabilities on competing operating systems. No doubt malicious actors have not waited to take the same tool at their advantage to unleash novel capabilities to their arsenal, beyond the malicious generative tools analyzed by the community: WormGPT\footnote{https://flowgpt.com/p/wormgpt-6}, DarkBERT \cite{jin2023darkbert}, FraudGPT \cite{falade2023decoding}.
In the domain of autonomous offensive cybersecurity operations, the probability and likely impact of proliferation of agentic AI frameworks are high. Understanding their mechanism to leverage these tools for defensive operations, and for being able to anticipate their malicious exploitation, is therefore an urgent requirement for the community.







We therefore propose the RedTeamLLM model to the community, as a proof-of-concept of the offensive capabilities of Agentic AI. The model encompasses automation, genericity and memory support. It also defines the principles of dynamic plan correction and context window contraint mitigation, as well as a strict security model to avoid abuse of the system. The evaluations demonstrate the strong competitivity of the model wrt. state-of-the-art competitors, as well as the necessary contribution of its summarizer, reasoning and act components. In particular, RedTeamLLM exhibit a significant improvement in automation capability against PenTestGPT \cite{deng2024pentestgpt}, which still show restricted capacity.

The remainder of this paper is organised as follows: Section \ref{sec:soa} presents the state of the art. Section \ref{sec:req} defines the requirements, and section \ref{sec:model} present the RedTeamLLM model for agentic-AI based offensive cybersecurity operations. Section \ref{sec:implem} presents the implementation and section \ref{sec:eval} the evaluation of the model.
Section \ref{sec:concl} concludes this work.

\section{State of the Art}
\label{sec:soa}

The advent, under the form of LLMs, of computing processes capable of generating structured output beyond existing text, is a key driver for a renewed development of agent-based models, with so-called `agentic AI' models \cite{shavit2023practices}, which are able both to devise technical processes and technically correct pieces of code. These novel kind of agents support multiple, complex and dynamic goals and can operated in dynamic environments while taking a rich context into account \cite{acharya2025agentic}. They thus open novel challenges and opportunity, both as generic problem-solving agents and for highly complex and technical environment like cybersecurity operations.

\subsection{Research challenges for Agentic AI}

The four main challenges in Agentic AI are: analysis, reliability, human factor, and production. These challenges can be mapped to the taxonomy of prompt engineering techniques by \cite{sahoo2024systematic}: \textit{Analysis:} Reasoning and Logic, knowledge-based reasoning and generation, meta-cognition and self-reflection; \textit{Reliability:} reduce hallucination, fine-tuning and optimisation, improving consistency and coherence, efficiency; \textit{Human factor:} user interaction, understanding user intent, managing emotion and tones; \textit{production:} code generation and execution.

The first issue for supporting reasoning and logic is the capability to address complex tasks, to decompose them and to handle each individual step. The first such model, chain-of-thought (CoT), is capable of structured reasoning through step-by-step processing and proves to be competitive for math benchmarks and common sense reasoning benchmarks \cite{wei2022chain}. Automatic chain-of-thought (Auto-CoT) automatize the generation of CoTs by generating alternative questions and multiple alternative reasoning for each to consolidate a final set of demonstrations \cite{zhang2022automatic}. Trees-of-thought (ToT) handles a tree structure of intermediate analysis steps, and performs evaluation of the progress towards the solution \cite{yao2023tree} through breadth-first or depth-first tree search strategies. This approach enables to revert to previous nodes when an intermediate analysis is erroneous. Self consistency is an approach for the evaluation of reasoning chains for supporting more complex problems through the sampling and comparative 1evaluation of alternative solutions \cite{wang2022self}.

Text generated by LLM is intrinsically a statistical approximation of a possible answer: as such, it requires 1) a rigorous process to reduce the approximation error below usability threshold, and 2) systematic control by a human operator. The usability threshold can be expressed in term of veracity, for instance in the domain of news.\footnote{https://www.cjr.org/tow\_center/we-compared-eight-ai-search-engines-theyre-all-bad-at-citing-news.php}.
 For code generation, it matches code that is both correct and effective, \textit{i.e.} that compiles and run, and that perform expected operation. Usable technical processes, like in red team operations, are defined by reasoning and logic capability.
reducing hallucination: Retrieval Augmented Generation (RAG) for enriching prompt context with external, up-to-date knowledge \cite{lewis2020retrieval}; REact prompting for concurrent actions and updatable action plans, with reasoning traces \cite{yao2023react}.

One key issue for red teaming tasks is the capability to produce fine-tuned, system-specific code for highly precise task. Whereas the capability of LLMs to generate basic code in a broad scope of languages is well recognized \cite{li2024evocodebench}, the support of complex algorithms and target-dependent scripts is still in its infancy.
In particular, the articulation between textual, unprecise and informal reasoning and lines of code must solve the conceptual gap between the textual analysis and the executable levels. Structured Chain-of-Thought \cite{li2025structured} closes this gap by enforcing a strong control loop structure (if-then; while; for) at the textual level, which can then be implemented through focused code generation. Programatically handling numeric and symbolic reasoning, as well as equation resolution, requires a binding with external tools, such as specified by Program-of-Thought (PoT) \cite{bi2024program} or Chain-of-Code (CoC) \cite{li2023chain} prompting models. However, these features are not required in the case of read teaming tasks.

\subsection{Cognitive Architectures}

Three main architectures implement the Agentic AI approach: ReAct (Reason and Act), ADaPT (As needed Decomposition and Planning) and P\&E (Plan and Execute).

\begin{figure}[ht]
    \centering
    \includegraphics[width=0.5\textwidth]{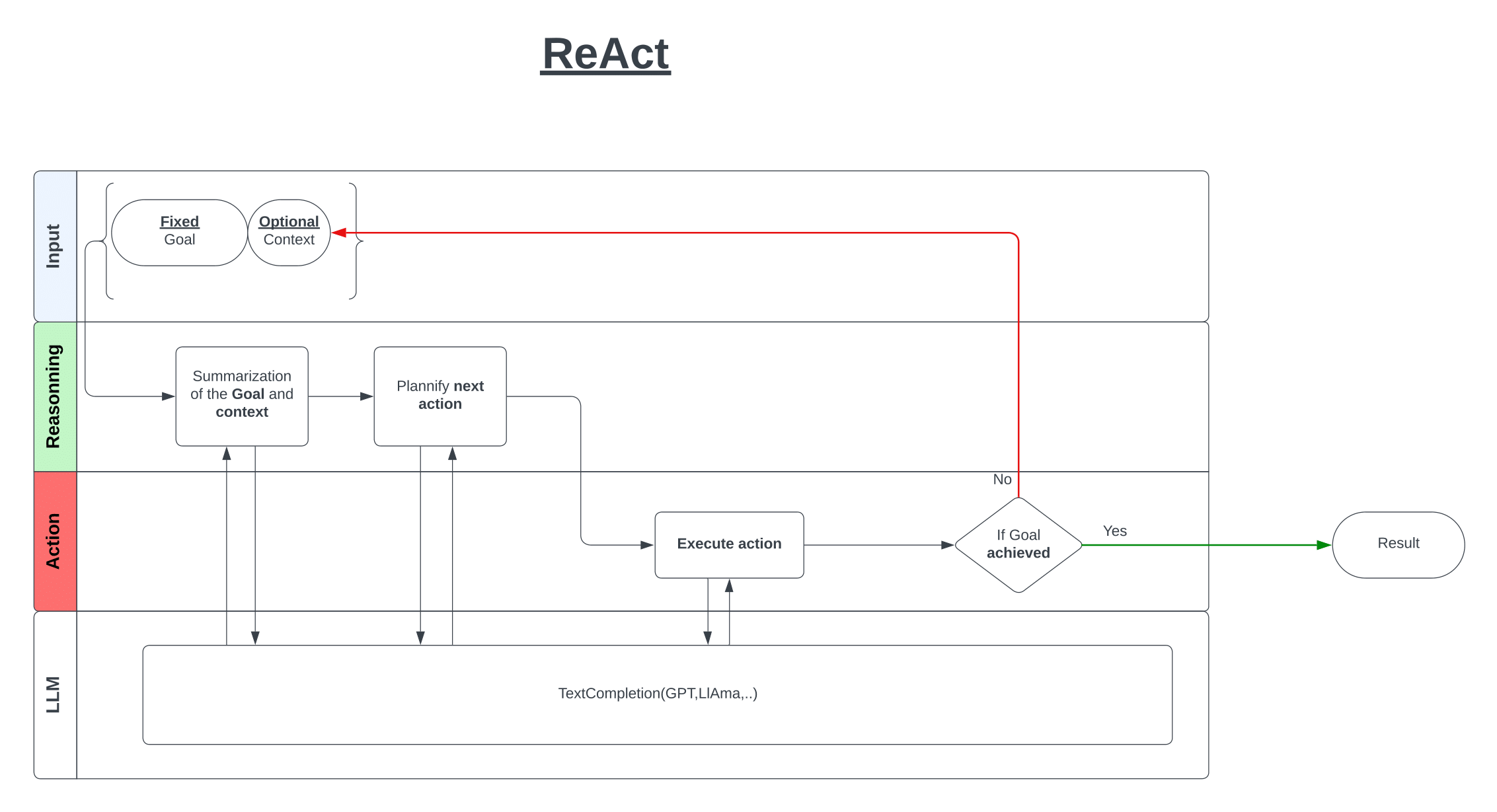} 
    \caption{Process diagram of ReAct}
    \label{fig:ReAct_Diagram} 
\end{figure}

ReAct\cite{yao2023react} first reasons about the analysis strategy, then rolls out this strategy. It performs multiple rounds of reasoning and acting, executing one action at each round then collecting observation. This enables a strong reduction of the error margin. As shown in Figure \ref{fig:ReAct_Diagram}, ReAct input is built with an explicit objective and an optional context. Reasoning then summarizes the goal and context and plan next action, each through a call to an LLM agent. The selected action is then executed, again based on an LLM call. If the analysis is not completed, the pipeline returns to the goal definition step, with a given subgoal. If the goal is achieved, the pipeline terminates. The main limits of this architecture, whether it is used with prompting or with complex pipelines, is the absence of memory, which requires each prompt to embed all context and knowledge about previous analysis steps. Since the context windows of current LLMs are strongly limited, information start being ignored as the context and history start exceeding the context window's limit,  which can lead to reduced performance and inaccurate outputs.

ADaPT \cite{prasad2023adapt} takes a greedy approach to decomposition: it keeps decomposing the task until it reaches subtasks that can be executed, through recursive decomposition which avoids a saturation of agent capability. The decomposition stops either when a task can be executed directly, or when a max depth is reached. Unlike ReAct and P\&E, ADaPT can't be a prompting method as it is based on recursion.
ADaPT completely solves the problem of context window size restriction, by decomposing as much as needed. Execution of leaves are then be carried out independently. However, many complications come along the way: plan correction (if a task fails completely, how can we correct the rest of the plan ?) and new discoveries (the agent might stumble upon information that can lead to a complete change of plan), in particular, are not supported.

P\&E \cite{sun2023pearl} aims to decompose a task into multiple subtasks that are executed independently from one another. This architecture defines first solutions to ReAct's weak points, by decomposing a task and isolating the subtask's execution. Prompt's length is thus minimized, which slows down the consumption of the context window capacity. Task execution becomes more efficient. However, one key issue remains: context window is eventually reached; and a new one is introduced: error handling, since, on a subtask's failure, the whole execution fails.

\subsection{Agentic AI and cybersecurity}

Recent offensive-security agents all converge on a narrow design spectrum: a frontier LLM in a ReAct-style loop that plans, executes a single tool call, observes, then repeats \cite{heckel2024countering} — yet none of them store or revise a global plan the way ADaPT or other deliberative-memory systems do. AutoAttacker couples ReAct with an episodic “Experience Manager,” but that memory is consulted only to validate the current action rather than to update or back-track the plan itself \cite{xu2024autoattacker}. LLM-Directed Agent preserves the classic four-stage ReAct chain (NLTG → CFG → CG → NLTP) and likewise discards alternative branches once the CFG selects one \cite{laney2024llm}.  One-Day  Vulnerabilities'  Exploit \cite{fang2024llm} and Hack-Websites \cite{fang2024llm-websites}  expose different toolsets to the same ReAct controller, and performance collapses as soon as GPT-4 is replaced by weaker models. CyberSecEval 3 uses an even leaner single-prompt ReAct wrapper to probe Llama-3 and contemporaries, finding that all models stall long before complex exploitation \cite{wan2024cyberseceval}. HackSynth strips the pattern down to just a Planner and a Summariser —- still a think-then-act loop—and shows that temperature and context-window size, not architectural novelty, dominate success rates \cite{muzsai2024hacksynth}. The sole departure from ReAct is PenTestAgent, which hard-codes a pentesting workflow (Reconnaissance → Search → Planning → Execution) without agentic recursion \cite{shen2024pentestagent}, and PenTestGPT, whose Plan-and-Execute modules shuffle intermediate results between Reasoning, Generation and Parsing stages but never revisit earlier strategies once execution starts \cite{deng2024pentestgpt}. Although defensive models exhibit promising properties \cite{ismail2025toward}, the exploitation of Agentic AI for malicious operations is a key concern to the community \cite{malatji2024artificial}.
Across current systems, memory is used only as a scratch-pad for latest observations; none implement hierarchical plan refinement, long-horizon memory, or roll-back of faulty plans.


\section{Requirements}
\label{sec:req}

In this section we explicit the specific challenges of agentic AI offensive cybersecurity operations.We address context window's limit, continuous improvement, genericity and automation.
One major issue of  LLM agent-based systems is their limited context window. Complex tasks usually require many iterations between the agent and a changing environment especially using ReAct, so tracking what has happened is essential for high-quality results. A common way to address this challenge is recursive planning \cite{prasad2023adapt}, in which a task is broken down into many subtasks that are executed individually; each subtask then passes the key points of its outcome to the next ones. A difficulty arises when a subtask fails, potentially blocking the subtasks that follow. To prevent this, a plan-correction mechanism \cite{wang2024tdag} is applied: whenever a subtask fails, the overall plan is adjusted so execution can proceed smoothly.
These two techniques are crucial for building a high-performance agent, but further refinements are still possible. Repeating the same mistakes on every run wastes time, money, and computation. Introducing a memory manager during task planning lets the agent avoid exploratory paths that have already failed.
Moreover, genericity is essential. Allowing the agent full freedom to choose its own tools and techniques fosters creativity and broadens its capabilities beyond a fixed toolset. In our case, the agent has unrestricted execution privileges through root access to a terminal.
Finally a key part to consider is automation; refining an agent system is important but not useful unless the whole process is automated, not requiring human interaction during the process.
Thus, integrating a tool call of an interactive terminal access within this context is rudimentary.

Consolidated requirements for our penetration-testing agent are thus:



\begin{enumerate}
    \item \textbf{Dynamic Plan Correction} --- Handling subtask or action failures without halting the entire workflow \cite{wang2025tdag}.
    \item \textbf{Memory Management} --- Managing large amounts of contextual data in long-running tasks, which enables continuous self-improvement.
    \item \textbf{Context Window Constraints mitigation} --- Preventing critical information loss due to an LLM’s limited prompt size \cite{yao2023react}.
    \item \textbf{Generality vs.\ Specialization} --- Balancing the need for specialized pentesting tools with broader adaptability.
    \item \textbf{Automation} ---  Automating the interaction of the agent with its designated environment; in our case a terminal. 

\end{enumerate}

\section{RedTeamLLM}
\label{sec:model}

In this section, we propose a novel architecture, supported features and related memory management mechanism for an offensive cybersecurity agentic model. Given the high capability and autonomy of the RedTeamLLM model, a robust security model is also required.
\subsection{The Architecture}

The architecture of RedTeamLLM is composed of seven components: \textbf{Launcher}, \textbf{RedTeamAgent}, \textbf{Memory Manager}, \textbf{ADaPT Enhanced}, \textbf{Plan Corrector}, \textbf{ReAct}, and \textbf{Planner}. On a run, the Launcher retrieves the input task and gives it to the \textbf{RedTeamAgent} while acting as the user interface (showing number of tasks running, memory access, failed and successful tasks, and allowing intervention in a task’s operation, e.g., stopping it or modifying its plan). Upon receiving the task, the \textbf{RedTeamAgent} has two objectives: pass it to \textbf{ADaPT Enhanced} and await a tree structure representing the full agent execution, then save that structure to the \textbf{Memory Manager}. \textbf{The Memory Manager}, which is the storage area for operation's knowledge, embeds and stores each node’s description from a task tree in a database, thus providing full access to previous task structures and dependencies. \textbf{ADaPT Enhanced} then takes that task, passes it to the \textbf{Planner} (which returns a tree of subtasks) and traverses it to execute leaves and pass results to siblings. The \textbf{Plan Corrector} can then adjust the plan and resume execution on any failure. All leaf executions are performed by the \textbf{ReAct} component, which carries out multiple rounds of reasoning, execution, and observations with terminal access. 

\begin{figure}[ht]
    \centering
    \includegraphics[width=0.5\textwidth]{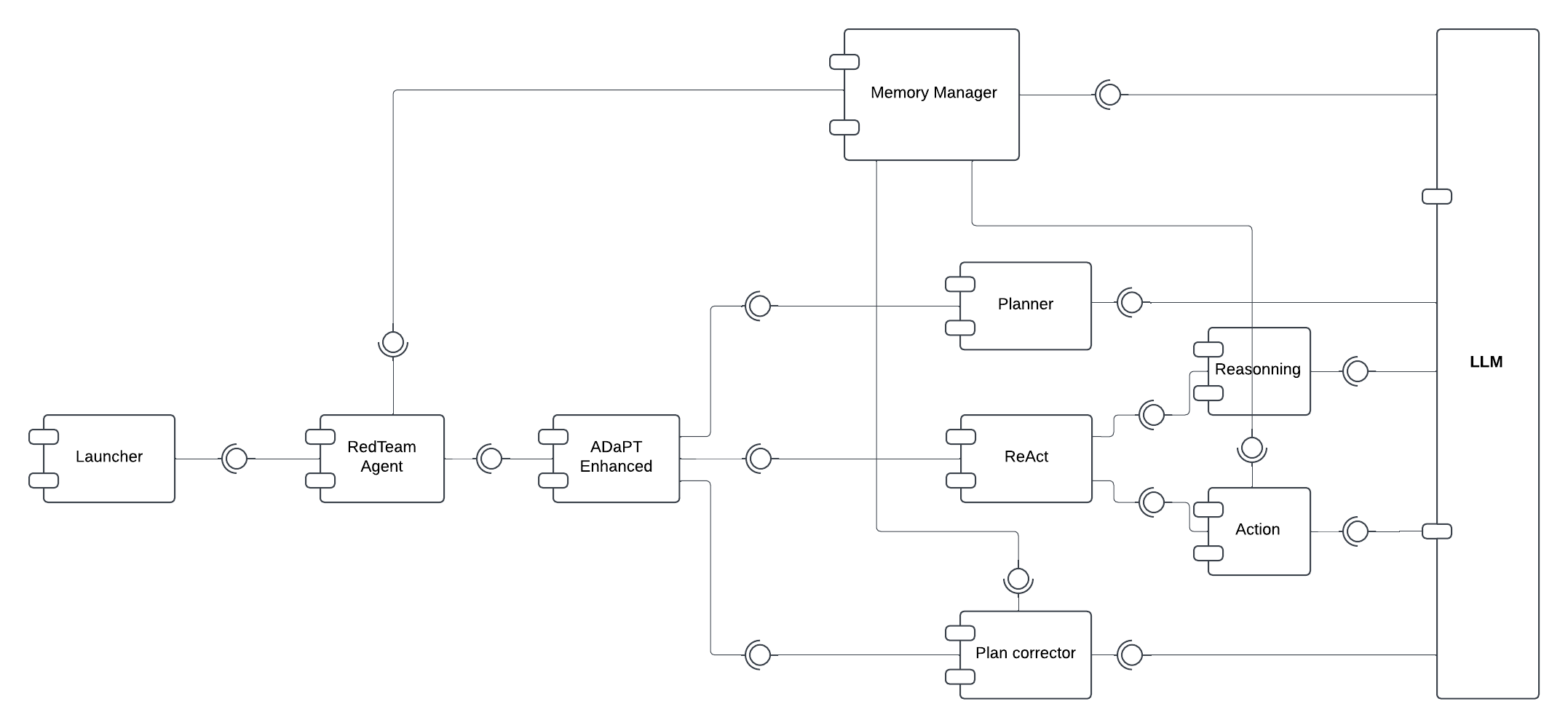} 
    \caption{Software Architecture for Red Team LLM Model}
    \label{fig:arch}
\end{figure}

\subsection{Features}
To support autonomous offensive operations, the proposed model must address many challenges and effectively meet essential requirements. Thus here are the principal features:

\begin{itemize}
    \item To address \textbf{context window's limit} the model needs to decompose a task recursively as much as needed. This is accomplished by the \textbf{ADaPT} component.
    
    \item With subtasks comes dependencies that need monitoring to avoid fatal execution failure on error. Here comes the \textbf{plan corrector} that has the ability to modify a task's accordingly to lattest outcomes.
    
    \item In order to support continuous improvement of the model capabilities, the \textbf{Memory Management} comes to improve planning over time by storing past execution in a tree-like way.
    
    \item Finally the model needs to be generic and avoid restriction to cover a wider range of tasks and not be limited to a set of tools. Here comes \textbf{ReAct} with full terminal access.This allows full automation, having full control and autonomy on the task it is executing.
\end{itemize}

\subsection{Memory management}
Memory management is an essential part of the model to be implemented. In all other competing models, memory is just used at the execution part to retrieve already executed commands for a similar task.
In our case, memory is used at a higher phase in which the agent decides how to create the execution plan. In fact, at the end of each execution the traces of the whole process are stored in form of a tree that is saved using task's description embedding. Thus, at every decomposition, the planner queries the saved node's hence having access to their success/failure reason, sub-tasks and detailed execution. This technique helps the agent improve over time and especially when re executing a task where he eventually narrows all the possibilities to the right path. This way the \textbf{RedTeamLLM model}, improves over time and has more chances to complete a task over multiple rounds of execution.
\begin{figure}[ht]
    \centering
    \includegraphics[width=0.5\textwidth]{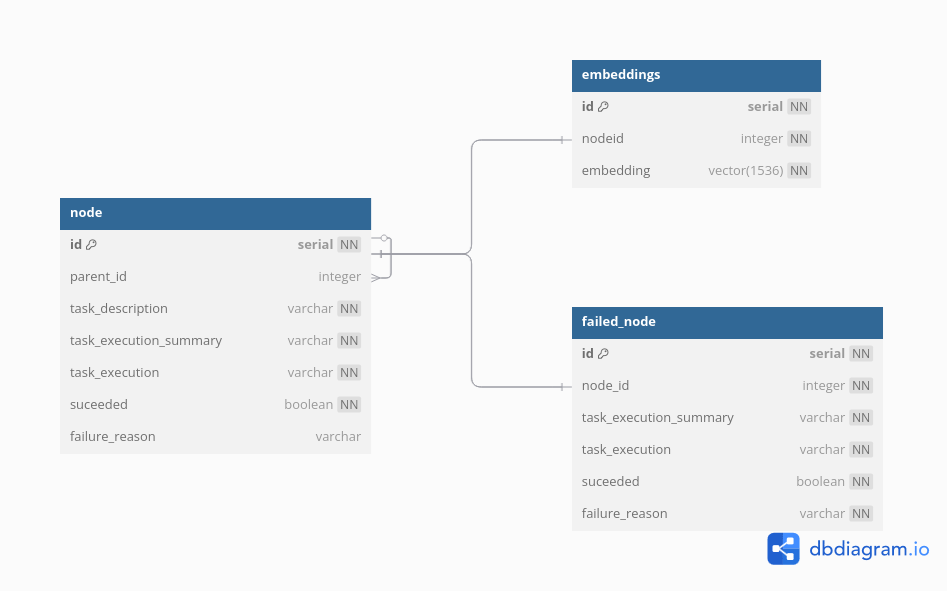} 
    \caption{Database schema for Memory management Model}
    \label{fig:memory_management_db_schema}
\end{figure}

\subsection{The Security Model}

\begin{figure}[ht]
    \centering
    \includegraphics[width=0.5\textwidth]{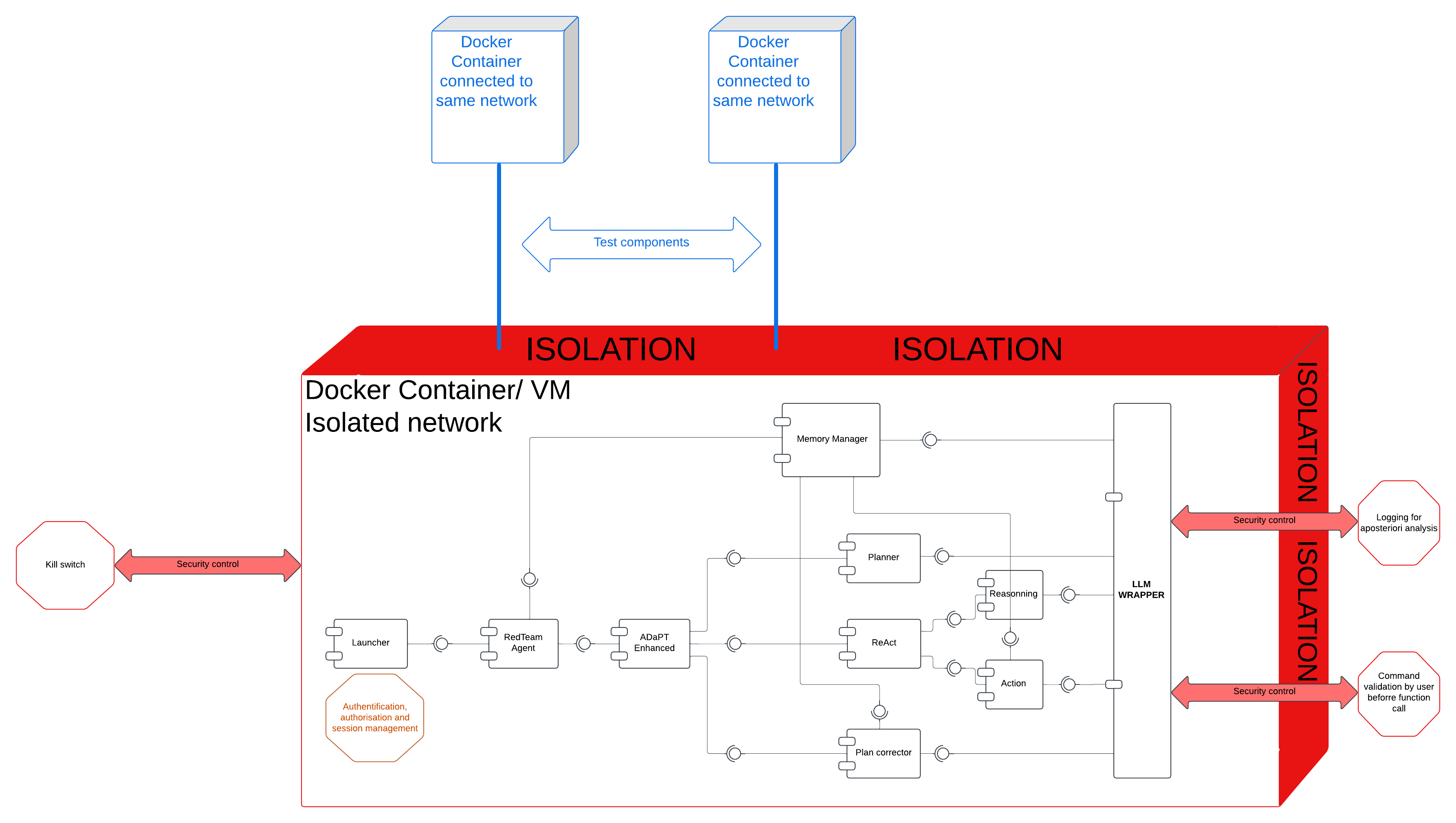} 
    \caption{Security layers wrapping the LLM agent}
    \label{fig:secu_model}
\end{figure}

\begin{figure}[ht]
    \centering
    \includegraphics[width=0.5\textwidth]{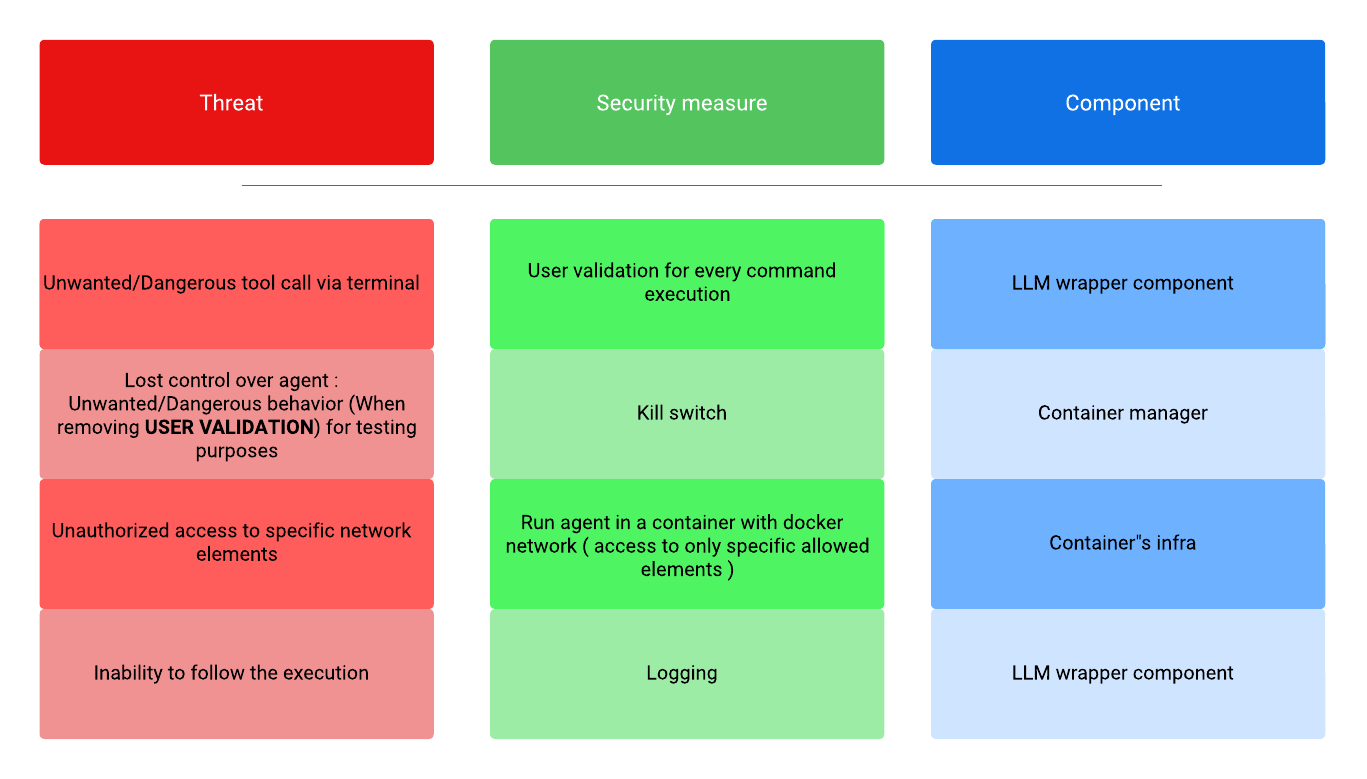} 
    \caption{Security challenges and how RedTeamLLM address them}
    \label{fig:security_challenges}
\end{figure}

The Red-Team LLM architecture supports a powerful autonomous process for pen-testing, including error recovery when the process meets dead-ends, and automation of offensive action. The architecture is thus exposed to two main threat families: hijacking of the execution process, on the one hand, and inversion of dependency from the LLM agents towards the framework, on the other hand. A strong security model is thus required to address the key vulnerabilities of agentic AI models: attack surface expansion, data manipulation and prompt injection, API usage and sensitive data exposure \cite{khan2024security}. Its five key components, shown in Figure \ref{fig:secu_model} are: 1) a dedicated authentication, authorization and session management module, 2) network and system isolation of the runtime environment, 3) systematic command validation by the user before any offensive action, 4) logging in append-only mode for a posteriori analysis and 5) a kill switch to shut the platform down. The threats related to containment and inversion of dependency are shown in table \ref{fig:security_challenges}. Isolation prevents unauthorized access to network entities or configurations, and to system capabilities. Command validation by the user ensures the alignment between the ongoing security task and performed operations, and prevent accidental calls to unwanted or dangerous tools upon proposal by the agent. Following and, when necessary, reconstructing the execution track is supported by the logging facility. To enhance the reaction capability and to pave the way to greater autonomy of the framework, a kill switch is set up to immediately halt any agent over which the supervision, or the control over actual operations, would have been weakened or lost.

The LLM itself is used in its default configuration, and with a benevolent user that have not intend to abuse it. Consequently, typical threats like prompt injection attacks \cite{labunetsfun} or app store abuses \cite{hou2024security} are not relevant to RefTeam LLM.

\section{Implementation}
\label{sec:implem}

The proof of concept for the RedTeamLLM model, that we evaluate in following section, entails the \textbf{ReAct} component for task execution. The current state of the implementation also covers \textbf{ADaPT} for recursive planning, \textbf{Memory management} for continuous improvement, and \textbf{Plan correction} to support operation continuity after task failure. However, these are less mature, and not evaluated here. RedTeamLLM and related tests are avalaible for the community\footnote{https://github.com/lre-security-systems-team/redteamllm}.

The evaluation is tested on a docker container over a Thinkpad e14 gen 5 with 16GB of RAM ddr4 /I513420h processor, and uses OpenAI's API with GPT4-o.

\subsection{Three-Step Pipeline}

The ReadTeamLLM implementation uses a three-step pipeline, each step handled by a separate LLM session:

\paragraph{1. Reasoning}
Before executing any action, the agent reasons about the next steps. Reasoning occurs in an isolated LLM session which elicits an explicit output of its process, detailed steps, and a plan. When the user provides the task definition to the model, it is forwarded to the reasoning component; its output is then passed to the Act component. After each tool call, the executed command and its output are fed back to the reasoning component to generate further analysis.

\paragraph{2. Act}
The output of Reasoning is treated as an assistant message by the Act session, which enforces adherence to the plan and reduces the model's inclination to interrupt execution with additional reasoning or safety checks. This setup allows the LLM to focus solely on executing the recommended action.
For tool execution, the LLM session has full access to a quasi-interactive, root-privileged Linux terminal. A current challenge is determining when a process requires input; we address this using \texttt{strace}, but it is not perfectly precise because some processes read from multiple file descriptors, not only \texttt{stdin}.
After each tool execution, if the output is too long, it is passed to a summarizer to avoid exceeding the context window.

\paragraph{3. Summarizer}
The summarizer is a stateless LLM session: for each request, it summarizes the given command’s output. Because this session does not maintain context about the agent’s overall goal, it sometimes omits important information. We plan to address this limitation in future work.

\subsection{Sample Run}

A sample run proceeds as follows:

\begin{enumerate}
\item A task is given to the agent (e.g., Obtain root access to the machine with IP \texttt{x.x.x.x}'').
  \item The task is forwarded to the reasoning session as a user message.
  \item The reasoner generates a result, which is provided as an assistant message to the acting session.
  \item The act session recommends a tool call (e.g., \texttt{nmap} or \texttt{sqlmap}).
  \item After execution, if the command output is lengthy, it is summarized and sent back to the reasoner as a user message.
  \item The reasoner produces further thoughts, and the loop continues until the reasoner stops recommending actions.
  \item At that point, the system prompts the user for input (e.g., Continue'' or a new task).
\end{enumerate}

\section{Evaluation}
\label{sec:eval}

The evaluation is performed in three steps: a qualitative evaluation of the RedTeamLLM capability to autonomously perform offensive operations; a comparative study between the cognitive mechanisms involved in these operations; an ablation study focused on the evaluation of the impact of the presence, or absence, of the reasoning capability.

\subsection{Use cases}

The choice of the benchmark to evaluate the RedTeamAgent is based on two factor: reproducibility and variability. We therefore selected 5 use cases: Sar, CewiKid, Victim1, WestWild, CTF4, from VULNHUB repository, which cover a broad range of technical difficulties and various security techniques, are easily deployable, and support reproducible executions. 
The objective of this work is focused on creating a proof on concept for ReadTeam LLM model, with the evaluation of cognitive operations: summarize, reason, act; and with a processing engine restricted to REaCT component. The 5 selected use cases are embedded in virtual machines from the easy category. This selections also allows us to compare our results since TAPT Benchmark \cite{isozaki2024towards} tested PentestGPT \cite{deng2024pentestgpt} on the same target VMs.

RedTeamLLM proves to be competitive for the target use cases, and surpasses PentestGPT on almost all the VM’s when using GPT4-o.
The decisive factors for these performance are the following ones.
First is reasoning, the difference without this step is really important. The agent used block more on same thoughts and doesn't keep a stable execution plan. Launching the agent multiple times, he sometimes completely changes strategy. Having an important amount of tokens dedicated to strategy, output analysis and reasoning help the agent to stay on track. Regularly, without reasoning the agent stops what he’s doing to ask for permission. Additionally, giving complete control over a terminal not giving a limited set of tools to the agent, helps with his creativity; being able to chose whatever path to take in order to achieve his goal. Sometime a specific version of a program isn't sufficient so he installs another one, sometime he launches scripts, sometimes he save operation information in a file. Moreover, the fact that he is directly executing the commands himself saves token on other topics. 
Finally automation is a key part of the agent, which enables longer and more complex automation without the need for manual supervision. 


\subsection{Cognitive steps}

\begin{figure}[ht]
    \centering
    \includegraphics[width=0.5\textwidth]{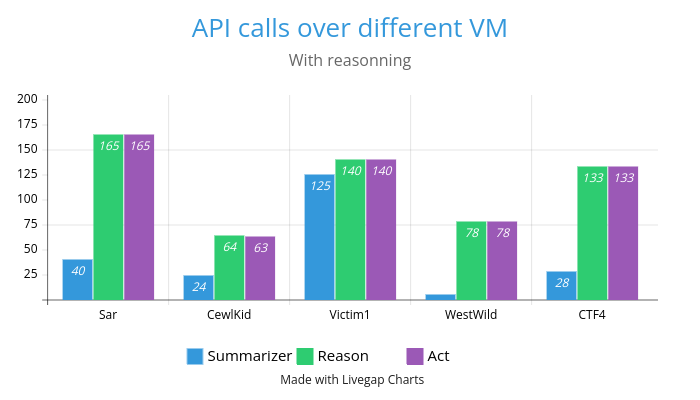} 
    \caption{Number of API calls in Summarizer, Reason, Act steps for the 5 use cases}
    \label{fig:API_Calls_reason}
\end{figure}

The RedTeamLLM implementation evaluated in this work is built around the ReACT analysis component. It entails 3 LLM session, \textit{i.e.} 3 interaction dialogs built by assistant and user messages: 3) the summarizer that summarizes command outputs; 2) the reasoning component that reasons over tasks and their outputs, and 3) the Act component that execute the tasks.
Figure \ref{fig:API_Calls_reason} shows the total number of API calls
for each component, over the different use cases after 10 tests on each VM.
The Summarizer typically consumes between 9,5\% (CTF4) and 15,9\% (Cewlkid) of API call tokens, with a low at 3,1\% for the WestWild use case and a peek at 30,9\% for the Victim1 use case. This peek enables a strong reduction of the required tool calls (See Fig. \ref{fig:Tool_Calls}). Reason and Act processes perform a very similar number of API calls.

RedTeamLLM outperforms PenTestGPT in 3 use cases out of 5: \textit{wrt.} the use case write-up, it completes 33\% more steps than PentestGPT-Llama (4 successful CTF levels vs. 3) and 300\% more than PentestGPT4-o (4 vs. 1) for Victim1 use case, 33\% more steps than PentestGPT4-o or PentestGPT-Llama (4 vs. 3) for WestWild use case, 75\% more than PentestGPT4-o (3.5 vs. 2) and 250\% than PentestGPT-Llama (3.5 vs. 1) for CTF4. PenttestGPT-Llama outperforms RedTeamLLM for Sar by 17\% (7 vs. 6) and by 100\% (4 vs. 2) for CewiKid use case, while PentestGPT4-o is similar or weaker that RedteamLLM for these 2 test cases.

\subsection{Reasoning: a strong optimization lever}

\begin{figure}[ht]
    \centering
    \includegraphics[width=0.5\textwidth]{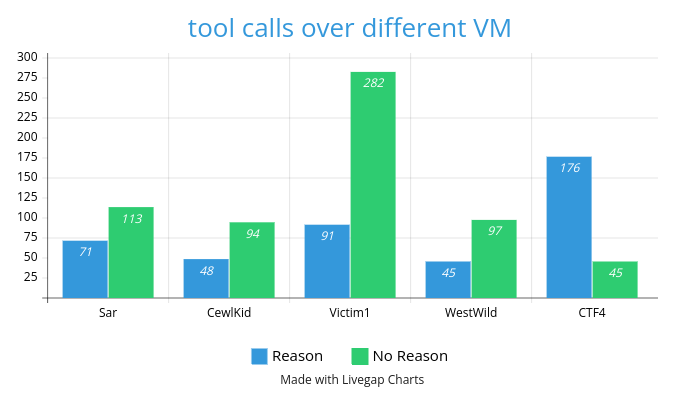} 
    \caption{Number of tool calls without and with reasoning for the 5 use cases}
    \label{fig:Tool_Calls}
\end{figure}

\begin{figure}[ht]
    \centering
    \includegraphics[width=0.5\textwidth]{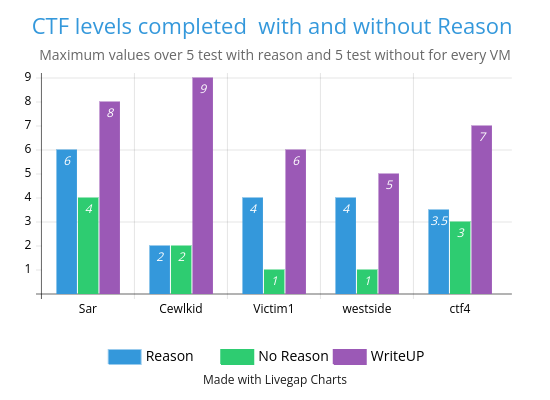} 
    \caption{CTF level completed by the RedTeamLLM framework without and with reasoning for the 5 use cases}
    \label{fig:ReAct_Ablation}
\end{figure}

The ablation study aims to evaluate the contribution of reasoning to the RedTeamLLM framework.
Figure \ref{fig:Tool_Calls} shows the number of tool calls without and with reasoning for the 5 use cases.
Every LLM session can have tool calls. A tool calls is a specific API response from an LLM session that triggers the use of provided tools (in our case a terminal). For example: when the agent executes a terminal command \texttt{ls}, that is a tool call response suggested by the LLM.
The total tool calls over the 5 vms with 10 tests on each VM is sumed up: 5 with reasoning, 5 without reasoning.
These are only the tool calls with the Act components only because this is where execution is performed. 
We can clearly see that the agent consumes significantly less tool calls with reasoning in 4 out of 5 use cases: the drop is tool calls range from 37\% (Sar) to 68\% (Victim1). Only for CTF4, the use of reasoning is bound with an increase of 291\% of tool calls, to support a slightly better achievement of the target operation (see Fig. \ref{fig:ReAct_Ablation}). In short, the agent performs more analysis before performing actions, and thus chooses better strategies to perform.

The degree of completion is computed for each use cases, using the write-up, which contains the listing of correct steps to complete the security challenge, as reference. Figure \ref{fig:ReAct_Ablation} shows these results. The write-up bar shows the total numbers of steps required 
to achieve the CTF (total of recon,general technique, exploit and privilege escalation).
The Reason and No Reason bars show how many steps the agent has completed for each use case with and without reasoning respectively.
The test process is similar to previous evaluation: RedTeamLLM handles 5 tests with reasoning and 5 tests without reasoning for every use case. The maximum number of steps achieved over the 5 runs is considered, \textit{i.e.} the better execution.
Reasoning improves the results in 4 cases out of 5. In two of these cases, the number of steps mastered pass from 1 to 4. A significant result of our experiments is that this improvement is coupled with a strong gain in efficiency wrt. to tool calls (see Fig \ref{fig:Tool_Calls}). In one case (Cewlkid), reasoning does not improve the offensive capability.

These results highlight the contribution of the reasoning step to security operation by RedTeamLLM model.

\section{Conclusions and Perspectives}
\label{sec:concl}

Beyond generative AI and now wide-spread Large Language Models (LLMs), Agentic AI is opening wide novel opportunities and threat to global security, and cybersecurity in particular. The objective of this work is to specify a reference model for agentic AI as applied to offensive cyber operations, so that the community can better understand these tools and their capability, leverage them for securing their information systems, and control this novel attack vector.

In this work, we define the key requirements for offensive agentic AI, propose a reference architecture model, and make a proof-of-concept of this architecture focused on iterative task analysis and execution through the ReACT component. The evaluation demonstrates that, though partial, our implementation beats state-of-the art competitors like PentestGPT in 60\% of the use cases. It also validate our hypothesis that reasoning is a key feature for agentic AI, since it enables a strong reduction of the necessary tool calls in 80\% of the use cases while improving offensive capabilities in 80 \% of the use cases. Interestingly enough, in 20\% of the use cases, it only supports reduction of tool calls, and thus process costs, and in 20\%, the gain in offensive capability requires a 4 times increase in tool calls. This proves that while RedTeamLLM improves both parameters in 60\% of cases, it is also efficient in dropping operation costs OR increasing operational capabilities in more complex tasks.

The key insight of this study is that leveraging the dual capability of LLMs to analyze and decompose processes, on the one hand, and to generate code for well-defined tasks, on the other, brings a radical improvement to automation and genericity of ReACT-based offensive cybersecurity frameworks. These first promising results pave the way to structuring the research effort in agentic AI for global security, in particular \textit{wrt.} methodologies for evaluation of cost and automation capabilities of these models. The evaluation of recursive planning, memory management and plan correction is also a necessity to better understand the underlying mechanics and capabilities of agentic models.






\bibliographystyle{named}
\bibliography{challita25ai4gs}



\end{document}